\documentclass{article}
\usepackage{amsmath}
\usepackage{amsfonts}
\usepackage{amssymb}
\usepackage{amscd}
\usepackage{graphicx}
\usepackage{color}

\newcommand{\beq}{\begin{equation}}
\newcommand{\eeq}{\end{equation}}

\begin{document}

\begin{center}
Scattering in a Euclidean formulation of relativistic quantum mechanics
\end{center}
\bigskip

\begin{center}
W. N. Polyzou, Gordon Aiello, Philip Kopp\\
Proceedings for 36th Annual International Symposium on Lattice Field Theory
\end{center}

\begin{itemize}
\item[] Abstract: A Euclidean formulation of relativistic quantum mechanics is
  discussed.  Representations of the Hilbert space inner product and
  Poincar\'e generators are all expressed in terms of Euclidean
  space-time variables.  The formulation does not require analytic
  continuation and can be used to directly calculate scattering
  observables. A toy model is used to demonstrate the feasibility of
  performing scattering calculations using the suggested computational methods.
\end{itemize}

\section{Euclidean relativistic quantum theory}

Relativistic treatments of quantum mechanics are needed to investigate
the dynamics and structure of physical systems on scales smaller than
the Compton wavelength of the constituent particles of the system.
This is particularly relevant for understanding strong-interaction
dynamics and structure at sub-nucleon distance scales.  While quantum
field theory provides an elegant and consistent treatment of
relativity and quantum mechanics, it is difficult formulate controlled
approximations that preserve relativistic invariance in kinematic
regions where the interactions are strong.  Finite degree of freedom
models that provide a consistent treatment of relativity and quantum
mechanics can also be formulated, however models that also satisfy
cluster properties normally require generating complicated
frame-dependent many-body interactions \cite{coester}.

The Euclidean axioms \cite{Osterwalder:1973dx} of quantum field theory have
the feature that the locality axiom
is logically
independent of the other axioms.  This suggests an alternate path to
formulate relativistic quantum models with a finite number of degrees
of freedom that retain all of the other desirable properties of the
field theory.  In addition, in the Euclidean case, the Green's
functions of the field theory are formally related to the Lagrangian of the
theory by a Euclidean path integral.  This can be used to include
dynamical constraints on models that are motivated by the local field
theory.

One of the surprising properties of the Euclidean construction is that
the quantum mechanical Hilbert space and all of the Poincar\'e
generators can be represented in a purely Euclidean framework without
performing an explicit analytic continuation in Euclidean time.  In
this work I discuss this construction along with how it can be used
to perform scattering calculations that are normally considered
un-natural in a Euclidean framework.

The dynamical input to the Euclidean formulation of relativistic quantum
mechanics is a collection of $N$-point Euclidean Green functions,
\beq
\{ G_{E:mk}(x_m , \cdots , x_1; y_1 , \cdots , y_k) \}   \qquad m+k=N
\label{eq.1}
\eeq
where $\{x_i\}$ are final Euclidean coordinates and $\{y_i\}$ are
initial Euclidean coordinates.  In a local theory there is only one
Green function for any pair $m+k=N$, however if locality is not
required there can be several for each $N$.  These Green functions
must be Euclidean invariant or covariant, symmetric or antisymmetric
with respect to interchange of the initial or final Euclidean
coordinates among themselves, satisfy cluster properties, be tempered
distributions, and satisfy reflection positivity.

The elements of a relativistic quantum theory are (1) a Hilbert space,
(2) a unitary representation of the Poincar\'e group satisfying
(3) cluster properties and (4) a spectral condition.  The Euclidean
reconstruction discussed below has all of these properties.

A dense set of elements of the Hilbert space are finite sequences of
functions of Euclidean space-time variables
\begin{equation}
\psi (x)  :=  ( \psi_1 (x_{11}), \psi_2 (x_{21},x_{22}),\cdots )
\label{eq.2}
\end{equation}
satisfying the positive relative-time support condition
\begin{equation}
\psi_n (x_{n1}, x_{n2},\cdots , x_{nn}) =0
\qquad
\mbox{unless} 
\qquad
0 < x_{n1}^0 <  x_{n2}^0  <\cdots <  x^0_{nn} .
\label{eq.3}
\end{equation}
The symmetry properties of the Green's functions imply that as long as
the Euclidean time support of these functions are satisfied for some
ordering,
the coordinates can be relabeled so (\ref{eq.3}) holds.
The physical Hilbert space inner product is
\[
\langle \psi \vert \phi \rangle_M = ({\theta} \psi ,{G_E} \phi)_E =
\sum_{nk} \int d^{4n}x d^{4k}y
\psi_n^* ( \theta  x_{n1}, \theta  x_{n2},\cdots ,
\theta x_{nn})
\times
\]
\begin{equation}
G_{E:nk}
( x_{nn}, \cdots , x_{1n};y_{1k} , \cdots , y_{kk})
\phi_k (y_{k1}, y_{k2},\cdots , y_{kk}) 
\label{eq.4}
\end{equation}
where $\theta$ is the Euclidean time reflection operator $
\theta x := \theta (\tau,\mathbf{x})= (-\tau,\mathbf{x})$.

Reflection positivity is the condition that $\langle \psi \vert \psi \rangle_M \geq 0$.  This ensures that (\ref{eq.4}) has the properties of a Hilbert space
inner product.

Relativistic invariance follows from the condition that the determinant
of the following $2\times 2$ matrices, 
\begin{equation}
X_m := \left (
\begin{array}{cc}
t+ z & x-iy\\
x+iy & t-z
\end{array}
\right )
\qquad
X_e := 
\left (
\begin{array}{cc}
i\tau + z & x-iy\\
x+iy & i\tau-z
\end{array}
\right ),
\label{eq.5}
\end{equation}
is preserved under the linear transformation
$X \to X' = A X B^t$ for $\det (A) = \det (B)=1.$
These transformations, which preserve the Euclidean and Lorentz line
elements
\begin{equation}
  \mbox{\bf det} (X_M) = t^2 -\mathbf{x}^2 \qquad
  \mbox{\bf det} (X_E) = - (\tau^2 +\mathbf{x}^2),
\label{eq.6}
\end{equation}
define complex Lorentz and complex orthogonal transformations.
It follows that the real Euclidean transformations form a subgroup of
the complex Poincar\'e group.  This relation be exploited to relate
generators of the 4-dimensional Euclidean group to generators of
the Poincar\'e group.

In the spinless case the relations leads to the following
representation of the Poincar\'e generators on the Euclidean representation of
the Hilbert space (on each component of $\psi$):
\begin{equation}
H \psi_n (x_{n1}, x_{n2},\cdots , x_{nn})
=
\sum_{k=1}^n {\partial \over \partial x_{nk}^0} \psi_n (x_{n1}, x_{n2},\cdots , x_{nn})
\label{eq.7}
\end{equation}
\begin{equation}
\mathbf{P}  \psi_n (x_{n1},x_{n2},\cdots , x_{nn})
=
-i \sum_{k=1}^n {\partial \over \partial \mathbf{x}_{nk}} \psi_n (x_{n1}, x_{n2},
\cdots ,x_{nn})
\label{eq.8}
\end{equation}
\begin{equation}
\mathbf{J} \psi_n ( x_{n1}, x_{n2},\cdots ,
x_{nn}) = -i \sum_{k=1}^n \mathbf{x}_{nk} \times {\partial \over
\partial \mathbf{x}_{nk}} \psi_n (x_{n1},
x_{n2},\cdots ,x_{nn})
\label{eq.9}
\end{equation}
\begin{equation}
\mathbf{K} \psi_n ( x_{n1}, x_{n2},\cdots ,
x_{nn}) = \sum_{k=1}^n (\mathbf{x}_{nk} {\partial \over
\partial {x}^0_{nk}}-  {x}^0_{nk}  {\partial \over
\partial \mathbf{x}_{nk}}) \psi_n ( x_{n1},
x_{n2},\cdots , x_{nn}),
\label{eq.10}
\end{equation}
which are the generators of time translation, space translation,
rotations and rotationless Lorentz boosts respectively.  These
generators satisfy the Poincar\'e commutation relations, and are formally
Hermitian with respect to the inner product (\ref{eq.4}).  In these
expressions all of the variables and derivatives are Euclidean.  If
the Green functions satisfy cluster properties then these generators
also satisfy cluster properties.  In addition, if the Green functions
satisfy reflection positivity, then it follows that the
Hamiltonian above satisfies a spectral condition.

\section{Scattering  theory}

Given a Hilbert space and a Hamiltonian satisfying cluster properties,
scattering observables can be defined and calculated using the same
methods that are used in non-relativistic scattering theory.  
In a quantum theory the $S$-matrix is 
the probability amplitude for scattering from an initial
state to a final state
\begin{equation}
S_{fi} := \langle \psi_+ \vert \psi_- \rangle .
\label{eq.11}
\end{equation}
The scattering state vectors can be expressed 
in terms of free state vectors $\vert \psi_{0\pm} \rangle$ that are
seen asymptotically in the detector after the collision or in the beam and
target before the collision
\begin{equation}
\vert \psi_{\pm} \rangle = \Omega_{\pm} \vert \psi_{0\pm} \rangle .
\label{eq.12}
\end{equation}
Wave operators $\Omega_{\pm}$ for multichannel scattering have the general
structure
\[
\Omega_{\pm} \vert \psi_{0\pm} \rangle =
\lim_{t \to \pm \infty}\sum
e^{iHt}\underbrace{\prod_n\vert \phi_n, \mathbf{p}_n, \mu_n \rangle}_{J}
\underbrace{e^{-ie_nt}}_{e^{-iH_0 t}} \underbrace{f_n(\mathbf{p}_n,\mu_n)}_{\vert \psi_0\pm\rangle} d\mathbf{p}_n  
\]
\begin{equation}
=: \lim_{t \to \pm \infty} e^{iHt}J e^{-iH_0t} \vert \psi_0\pm\rangle
\label{eq.13}
\end{equation}
where $\vert \phi_n,\mathbf{p}_n,\mu_n \rangle$ represents an elementary
or bound system with total momentum $\mathbf{p}_n$, magnetic quantum number
$\mu_n$, and energy $e_n$.  $f_n(\mathbf{p}_n,\mu_n)$ represents a localized
wave packet with the mean momentum of the particle or bound sub-system.
The mapping $J$ \cite{coester} above is called an injection operator.  It is a mapping from
a Hilbert space of scattering asymptotes to the physical Hilbert space.

Haag-Ruelle scattering is the field-theory version of the above.  In
the field theory case the operator $J$ is expressed as a suitably
symmetrized product of operators that create single-particle states out
of the vacuum.  While this requires solving the one-body problem,
the benefit is that the limits in (\ref{eq.13})
(\cite{haag}\cite{ruelle}\cite{jost}) are strong limits.

In the Euclidean case for a two-particle initial state a candidate
for the injection operator $J$ is \cite{gordon}
\[
J :
\langle x_1 \vert  \phi_1, \mathbf{p}_1 \rangle 
\langle x_2 \vert  \phi_2, \mathbf{p}_2 \rangle =
\]
\begin{equation}
{h_1 (\nabla_1^2)}\delta (x_1^0 - \tau_1 )
{h_2 (\nabla_2^2)}\delta (x_2^0 - \tau_2 )
{1 \over (2 \pi)^3} e^{i \mathbf{p}_1\cdot  \mathbf{x}_1
+ i \mathbf{p}_2\cdot  \mathbf{x}_2}
\label{eq.14}
\end{equation}
\begin{equation}
\tau_2 > \tau_1
\label{eq.15}
\end{equation}
where the Euclidean Laplacians, $\nabla_i^2$, are the mass squared operators
for each particle or subsystem, and the $h_i(m^2)$ are smooth functions
that are 1 when $m$ is the mass of the $i^{th}$ 
particle (or subsystem) and 0 on the rest of the mass spectrum.
The delta function in the Euclidean times ensures the time-support condition.

The reason that this is only a candidate is because
$h_2 (\nabla^2)$ is not analytic in $\nabla^2$, so it could
transform a wave functions satisfying the relative-time support
conditions to one that does not, leading to a range that is out of
Hilbert space.  This will not happen if polynomials in
$\nabla^2$ are complete in this space.  To establish this, note that
a sufficient condition for completeness is that the Stieltjes moments
\begin{equation}
\gamma_n := \int_0^\infty {e^{- \sqrt{m^2 + \mathbf{p}^2} \tau} 
\over 2 \sqrt{m^2 + \mathbf{p}^2}} \rho(m) 
m^{2n} dm 
\label{eq.16}
\end{equation}
where 
$\tau = \tau_1 + \tau_2 >0$
satisfy 
Carleman's condition \cite{carleman}
\begin{equation}
\sum_{n=0}^\infty  \vert \gamma_n \vert^{-{1\over 2n}} > \infty .
\label{eq.17}
\end{equation}
This will hold as long as the Lehmann weight $\rho(m)$ in (\ref{eq.16}) is
polynomially bounded \cite{gordon}.   This ensures that the Haag-Ruelle functions
$h(\nabla^2)$ can be approximated by polynomials.

A sufficient condition for the  convergence of the limit (\ref{eq.13})
that defines
the scattering wave operators is the Cook condition \cite{cook}, which in the
Euclidean representation for 2-2 scattering has the form
\begin{equation}
\int_a^\infty \Vert  (HJ - J H_0)
e^{\mp iH_0t}  \vert \psi_{0} \rangle
\Vert_M dt < \infty
\label{eq.18}
\end{equation}
where $a$ is a constant and 
\[
\Vert  (HJ - J H_0)
\Phi e^{\mp iH_0t} \vert \psi_{0} \rangle
\Vert_M^2 :=
\]
\begin{equation}
(\psi_0 e^{\pm iH_0t} (J^{\dagger} H - H_0 J^{\dagger})
\theta G_E (HJ - J H_0)
e^{\mp iH_0t} \vert \psi_{0} )_E .
\label{eq.19}
\end{equation}
The important observation is that because the injection operator
asymptotically projects on one-body states, the definitions imply that
the contribution to the integral (\ref{eq.19}) due to the disconnected parts
of the Green function vanishes \cite{polyzou}\cite{gordon}.
All that remains is the connected
part, which, if the spectrum has a mass gap, is expected to fall off
like $t^{-3}$ for large $t$.  This suggest that the scattering problem
is mathematically well defined in this Euclidean representation.

\section{Computational considerations}

There are a number of tricks that can facilitate the computation of
scattering observables in the Euclidean case \cite{kopp}.  The first is to use the
invariance principle \cite{simon} which implies 
\begin{equation}
\lim_{t \to \pm \infty} e^{iHt} J e^{-iH_0t } \vert \psi \rangle =
\lim_{t \to \pm \infty} e^{if(H)t} J e^{-if(H_0)t } \vert \psi \rangle 
\qquad \mbox{for} \qquad 
f(x) = -e^{-\beta x} .
\label{eq.20}
\end{equation}
This gives
\begin{equation}
\lim_{t \to \pm \infty} e^{iHt} J e^{-iH_0t } \vert \psi \rangle =
\lim_{n \to \infty} e^{\mp i ne^{-\beta H}} J e^{i \pm ne^{-\beta H_0}} \vert \psi \rangle .
\label{eq.21}
\end{equation}
Since $\sigma(e^{-\beta H}) \in [0,1]$, this means that 
$e^{\mp i ne^{-\beta H}}$ can be uniformly approximated by
a polynomial in $e^{-\beta H}$ where $\beta>0$ is a parameter
that can be adjusted for convergence. The inequality  
\begin{equation}
\vert e^{inx} - P(x) \vert < \epsilon \qquad x \in [0,1]
\label{eq.22}
\end{equation}
leads to the uniform operator inequality
\begin{equation}
\vert\Vert e^{ine^{-\beta H}} - P (e^{-\beta H}) \vert\Vert < \epsilon
\label{eq.23}
\end{equation}
with the same $\epsilon$ in (\ref{eq.22}) and (\ref{eq.23}). 
This is useful in the Euclidean case
because $e^{-n\beta H}$ simply translates the Euclidean time to the
right by $n\beta$.

This method can be used to calculate sharp-momentum transition
matrix elements.  The approximate expression has the form
\begin{equation}
\langle \mathbf{k}_f \vert T(E+i0) \vert \mathbf{k}_i \rangle
\approx
\langle \psi_{f0} \vert (J^{\dagger}H-H_0J^{\dagger}) P(e^{-\beta H}) J e^{i ne^{-\beta H_0}}\vert
\psi_i \rangle
\label{eq.24}
\end{equation}  
where the initial and final wave packets must be sufficiently narrow,
$\beta$ must be chosen based on the energy scale, $n$ must be
sufficiently large and the polynomial $P(x)$ must accurately approximate
$e^{2inx}$ on $[0,1]$.  These approximations must be done in the
proper order. (1) First choose a sufficiently narrow wave packet in
momentum space. (2) Choose $\beta$ based on the energy scale. (3)
For the choice of $\beta$ and
wave packet choose $n$ large enough for convergence. (4) Given the
$n$ from step 3 construct a polynomial approximation to $e^{2inx}$
for $x\in [0,1]$.

This computational method was tested for scattering from a separable
potential with the range of a pion exchange interaction and a strength
chosen to produce a bound state with the binding energy of a Deuteron:
\begin{equation}
\langle \mathbf{k}' \vert V \vert \mathbf{k} \rangle  
= -{\lambda \over
(m_\pi^2 +\mathbf{k}^{\prime 2})(m_\pi^2 +\mathbf{k}^2)}.
\end{equation}
While this is not a Euclidean calculation, it is exactly solvable and
provides a means to precisely test the use of the invariance
principle, narrow wave packets and the polynomial approximation to $e^{ine^{-\beta H}}$ in the
computation of sharp-momentum transition matrix elements.
In this calculation
the sharp-momentum transition matrix elements were extracted directly
from $S$-matrix elements rather than using (\ref{eq.24}).

The result of this exploration was that convergence was achieved for a
wide range of momenta, between 50 MeV and 2 GeV \cite{kopp}.  The parameters of
the approximations were chosen to get better than a 1\% error in the
scattering amplitude.  The largest source of error (by far) was the
wave packet width.  For wave packets chosen to give an error better than
1\% 
the $n$ values were a few hundred and the degree
of the polynomial was slightly higher.  The polynomials were
accurately approximated using a Chebyshev expansion.  The parameter
$\beta$ was chosen so $\beta E$ was a number of order unity.  The
results are shown in the table

\vbox{
\begin{center}
Table 1: Sharp-momentum transition matrix elements
\end{center}
\begin{center}
\begin{tabular}{llll}
\hline
$k0$ & Real T & Im T & \% error \\
\hline
0.05  &   2.18499e-1    &  -1.03160e+0	          &     0.0982   \\
0.1   &  -2.30337e-1    &  -4.09325e-1	          &     0.0956   \\
0.2   &  -1.01512e-1    &  -4.61420e-2	          &     0.0981   \\
0.3   &  -3.46973e-2    &  -6.97209e-3	          &     0.0966   \\
0.4   &  -1.39007e-2    &  -1.44974e-3	          &     0.0997   \\
0.5   &  -6.44255e-3    &  -3.86459e-4	          &     0.0986   \\
0.6   &  -3.34091e-3    &  -1.24434e-4	          &     0.0952   \\
0.7   &  -1.88847e-3    &  -4.63489e-5           &     0.0977   \\
0.8   &  -1.14188e-3    &  -1.93605e-5           &     0.0965   \\
0.9   &  -7.28609e-4    &  -8.86653e-6           &     0.0982   \\
1.0   &  -4.85708e-4    &  -4.37769e-6           &     0.0967   \\
1.1   &  -3.35731e-4    &  -2.30067e-6           &     0.0987   \\
1.2   &  -2.39235e-4    &  -1.27439e-6           &     0.0968   \\
1.3   &  -1.74947e-4    &  -7.38285e-7           &     0.0985   \\
1.4   &  -1.30818e-4    &  -4.44560e-7           &     0.0955   \\
1.5   &  -9.97346e-5    &  -2.76849e-7           &     0.0956   \\
1.6   &  -7.73390e-5    &  -1.77573e-7           &     0.0992   \\
1.7   &  -6.08794e-5    &  -1.16909e-7           &     0.0964   \\
1.8   &  -4.85672e-5    &  -7.87802e-8           &     0.0956   \\
1.9   &  -3.92110e-5    &  -5.42037e-8           &     0.0967   \\
2.0   &  -3.20000e-5    &  -3.80004e-8           &     0.0966   \\
\hline
\end{tabular}
\end{center}
}

These calculations support the possibility of directly performing
scattering calculations in a Euclidean representation.  The main
challenge for using this method is how to include dynamics.  The
physics is in the Euclidean Green functions; these have to be computed
or modeled.  Reflection positivity puts strong constraints on these
models.  A structure theorem for reflection positive distributions is
an important goal of this research program.  This is needed to
identify acceptable models that have Hilbert space inner products.
For a theory like QCD both cluster properties and reflection
positivity are limited to initial and final states being local color
singlets.  For this reason a formulation directly in terms of
gauge-invariant degrees of freedom is another important goal.

This work supported in part by the U.S. Department of Energy Office of Science, grant DESC0016457

\end{document}